\documentstyle[12pt,aasms4]{article}
\tighten

\def\lsim{\lower.5ex\hbox{$\; \buildrel < \over \sim \;$}}
\def\gsim{\lower.5ex\hbox{$\; \buildrel > \over \sim \;$}}
\def\lax    {\ifmmode{_<\atop^{\sim}}\else{${_<\atop^{\sim}}$}\fi}
\def\gax    {\ifmmode{_>\atop^{\sim}}\else{${_>\atop^{\sim}}$}\fi}
\def\etal{{\it et al.\/} }
\def\gtorder{\mathrel{\raise.3ex\hbox{$>$}\mkern-14mu
             \lower0.6ex\hbox{$\sim$}}}
\def\ltorder{\mathrel{\raise.3ex\hbox{$<$}\mkern-14mu
             \lower0.6ex\hbox{$\sim$}}}

\def\pmb#1{\setbox0=\hbox{#1}%
  \kern-0.015em\copy0\kern-\wd0
  \kern0.03em\copy0\kern-\wd0
  \kern-0.015em\raise0.0433em\box0 }

\begin{document}

\title{ $\delta-$invariant for Quasi-periodic Oscillations
and Physical Parameters of 4U 0614+09 binary}

\author{Lev Titarchuk}
\affil{NASA/ Goddard Space
Flight Center, Greenbelt MD 20771, and George Mason University/CSI, USA;
lev@lheapop.gsfc.nasa.gov}

\author{Vladimir Osherovich}
\affil{NASA/Goddard Space Flight Center/RITSS, Greenbelt MD 20771 USA;
vladimir@urap.gsfc.nasa.gov}

\vskip 0.5 truecm


\begin{abstract}
The recently formulated Two Oscillator (TO) model  interprets the lowest of 
the kilohertz frequencies of the twin peak quasi-periodic oscillations in X-ray 
binaries as the Keplerian frequency $\nu_{\rm K}$. The high twin 
frequency $\nu_h$ 
in this model holds the upper hybrid frequency relation to the rotational 
frequency of the neutron star's magnetosphere ${\bf\Omega}$:
$\nu_h^2=\nu_{\rm K}^2+4(\Omega/2\pi)^2$. 
The vector ${\bf\Omega}$ is assumed to have an angle $\delta$ with the normal 
to the disk. The first oscillator in the TO model allows one to interpret 
the horizontal branch observed below 100 Hz as the lower 
mode  of the Keplerian oscillator under the influence of 
the Coriolis force, with frequency $\nu_L$ being dependent on 
$\nu_h$, $\nu_{\rm K}$ and $\delta$. 
For some stars such as 4U 0614+09, Sco X-1 and 
4U 1702-42,  $\nu_h$, $\nu_{\rm K}$ and $\nu_L$ have been observed 
simultaneously providing the opportunity to check the central prediction 
of the TO model: the constancy of $\delta$ for a particular source. 
Given the considerable variation of  each of these three frequencies, 
the existence of an observational invariant with a clear physical 
interpretation as a global parameter of the neutron star magnetosphere  
is an important test of the TO model.
Using the results of recent observations of 4U 0614+09 
we verify the existence  of this invariant and determine 
the angle $\delta=15^o.6\pm 0.^o5$  
for this star. 
The second oscillator in the model deals with a
radial (presumably sound) oscillation  and diffuse process in the viscous layer
surrounding the neutron star. Our analysis of the viscous oscillation frequency
$\nu_V$  and the break frequency $\nu_b$ of the diffusion 
 shows that the spin value of the inner boundary of the transition 
layer for 4U 0614+09 is  at least two times more than values for 4U 1728-34 
and Sco X-1. 

\end{abstract}


\section{Introduction}

This {\it Letter} contains  further verification of 
the Two Oscillator (TO) model for kHz QPO's in NS binaries and the related 
low frequency features using the observational results  for 4U 0614+09 
from van Straaten et al. (2000) (hereafter VS00).
In a series of the papers (Osherovich \& Titarchuk 1999a, 
Titarchuk \& Osherovich 1999, Osherovich \& Titarchuk 1999b, Titarchuk, 
Osherovich \& Kuznetsov 1999, hereafter OT99a, TO99, OT99b and TOK 
respectively) the authors offered the TO model which  explained the 
variation of the  frequency difference between two kHz QPOs suggesting
that the upper one, namely $\nu_h$ is an upper hybrid frequency
of the Keplerian oscillator under the influence of the Coriolis force and
{\it the lower kHz QPO is the Keplerian frequency $\nu_K$}.
This identification is a principal difference between the TO model and 
other QPO models
presented so far (e.g. Stella \& Vietri 1998  Miller, Lamb, Cook 1998, 
Kaaret, Ford \& Chen 1997)  since in other models, the highest kHz QPO is 
identified as a Keplerian frequency. 

According to the TO model, the Keplerian oscillator
has two branches characterized by high frequency $\nu_h$ ($\sim$ 1 kHz)
 and by low frequency $\nu_L$ ($\sim 50-100$ Hz). The frequency $\nu_L$ depends
strongly on the angle $\delta$ between the normal to the neutron star
disk and $\bf{\Omega}-$the angular velocity of the magnetosphere surrounding
the neutron star. In the lower part of the QPO spectrum ($\sim$ 10 Hz),
the second oscillator of the TO model describes the physics of the viscous 
transition layer, namely, radial   
viscous oscillations with frequency
$\nu_{V}$ (previously called \lq\lq extra noise component") and
 the diffusive process 
in the transition region (the innermost part of the disk) 
which is characterized by the break frequency $\nu_b$. 
According to the TO model, all frequencies (namely $\nu_h$, $\nu_L$, $\nu_b$
and $\nu_{V}$) have specific dependences on $\nu_{\rm K}$.
Correlations of $\nu_b$ and $\nu_V$ with kHz frequencies  are fitted  in TO99 
by the theoretical curves using the dimensionless parameter, $a_K$: 
\begin{equation}
a_{\rm K}=m(x_0/3)^{3/2}(\nu_0/363~{\rm Hz})
\end{equation}
where normalized mass $m=M/M_{\odot}$, radius of the inner edge of disk 
$x_0=R_0/R_{\rm S}$ in Schwarzchild radius ($R_{\rm S}$ units)
and $\nu_0$ is the spin of the disk inner edge.  

Van Straaten et al. (2000)  have found
for 4U 0614+09 that dependences of $\nu_V$ and $\nu_b$ on $\nu_{\rm K}$
do not follow the theoretical curves which fit the data points for 4U1728-34
(TO99) and Sco X-1 (TOK).  
In this {\it Letter}, we revisit this issue and  show that the 
observational correlations for source 4U 0614+09 are related to the 
theoretical ones for the same parameter  $a_{\rm K}=1.03$ (as for 
4U 1728-34 and Sco X-1)  but with a spin of 
the inner edge of accretion disk higher than 363 Hz. Such a high spin leads 
to the high compactness of the central object because  
\begin{equation}
R_0=9~{\rm km}\times m^{1/3}
\left({{a_{\rm K}}\over{\nu_0/363~{\rm Hz}}}\right)^{2/3}.
\end{equation}
We check predictions of the 
TO model regarding  $\nu_V$ vs $\nu_{\rm K}$ dependence as well as relation 
between $\nu_b$ and $\nu_{\rm K}$ for all available observations (see \S3).

In OT99b, the authors demonstrated for the source 4U 1702-42 
that the inferred angle $\delta$, 
(see Eq 2, 5 there) 
\begin{equation}
\delta=\arcsin\left[(\nu_h^2-\nu_{\rm K}^2)^{-1/2}
(\nu_L\nu_h/\nu_{\rm K})\right]
\end{equation} 
 depends only, on  
$\nu_{\rm K}$, $\nu_h$ and  $\nu_L$ and stays the same  
  ($3.9^o\pm0.2^o$) over significant range of $\nu_{\rm K}$ (650-900 Hz). 
In Formula (3),  $\delta$ is assumed to be small.
A general formulation for any $\delta$ is presented, in
OT99a with conjecture that eventually $\delta$ as an invariant will be found 
for all 20 sources known to have QPOs.
Using  series of observations where $\nu_h$, $\nu_{\rm K}$, $\nu_L$
 are  detected simultaneously (van der Klis et al. 1997; 
Markwardt, Strohmayer \& Swank 1999; van Straaten et al. 2000)  we check that  
the angle $\delta$ is a true invariant  for  three specific sources:
Sco X-1, 4U 1702-42, 4U 0614+09 (see \S 2). 
 We will also present the
magnetospheric rotational profile for  4U 0614+09  inferred from 
VS00 data in \S 4.  
Discussion and summary follow in the last section.
\bigskip
\par
\noindent
\centerline{\bf 2. $\delta-$Invariant and Verification of Two-Oscillator Model}

We have used the frequencies measured for sources 4U 0614+09, Sco X-1 
and 4U 1702-42 where 
the low and high kHz QPO peaks $\nu_{\rm K}$
and $\nu_h$  and HBO frequencies $\nu_L$   are  
measured simultaneously.
The resulting values of $\delta$ calculated from Eq. (3) are shown in Figure 1. 
Indeed, for  each of these sources, the  $\delta-$values
 show little variation with $\nu_{\rm K}$, $\nu_h$, $\nu_L$.
They are $3^o.9\pm 0.2$ and $5^o.5\pm 0.5$ for 4U 1702-42 and Sco X-1 
respectively.
For the source 4U 0614+09 the
 point at $\nu_{\rm K}=821.6$ Hz is not included since 
the $\nu_{\rm K}$ peak is very broad with  FWHM= 205 Hz,
which is almost order of magnitude larger than that for any other points
for the same star according to Table 1 in VS00. 
The angle $\delta$ obtained  for the source  4U 0614+09 is 
$15^o.6\pm0^o.5$ and is significantly larger than for other stars.
The existence of   
the $\delta-$invariant  predicted by the TO model (OT99b) is a challenge for 
any other QPO model. It is important to note that   
{\it all frequencies included in this $\delta-$relation are observed 
frequencies and thus the relation Eq. (3) is a model independent invariant}.
\bigskip
\par
\noindent
\centerline{\bf 3. Break Frequency vs kHz QPO Frequency Correlations}

Further tests of the TO model can be done using a 
comparison of the observed correlation of break and  kHz  QPO frequencies
(VS00) with the theoretical dependences derived in Titarchuk, Lapidus \&
Muslimov (1998, hereafter TLM) and TO99.
In Fig. (2) (upper panel),  we present the theoretical curves calculated 
using Eq. (9) in TO99 for different values of $\nu_{0,363}=\nu_0/363$ Hz.
For the source 4U 0614+09 we found the Lebesgue's measure 
(for definition, see TOK) is better than 15\% for $a_{\rm K}=1.03$ and 
$\nu_{0,363}>2$  which is at least 3 times less than that for 
$\nu_{0,363}=1$ (see upper panel in Fig. 2). 
We  fitted  the data points by the curves with 
$\nu_{0,363}>3$ ($\nu_0> 1089$ Hz) 
and we have found that the quality of those fits is improved only slightly.
Thus it is difficult to get the constraints on $\nu_0$ better than 
$\nu_{0}\gtorder800$ Hz.  
Recently Bhattacharyya et al. (2000) using the full General Relativity 
technique argued that the NS in Cyg X-2 rapidly rotates with the frequencies 
of order of 1 kHz which is very close to that we obtained here for 4U 0614+09.

We approximate the solution of Eq (9) of 
TO99 by a polynomial for $a_{\rm K}=1.03$ and $\nu_{0,363}=1$, 
 namely,
\begin{equation}
\nu_{b}=C_b\cdot P_4(\nu_{\rm K})
=C_b(B_1\nu_{\rm K}+B_2\nu_{\rm K}^{2}+B_3\nu_{\rm K}^3+B_4\nu_{\rm K}^4)
\end{equation}
where $B_1=-1.68\times 10^{-4}$, 
$B_2=1.36 \times 10^{-5}$ Hz$^{-1}$, 
$B_3=-2.47\times10^{-8}$ Hz$^{-2}$, $B_4=4.45\times10^{-11}$ Hz$^{-3}$. 
We fit the observational break frequencies by  the theoretical 
curve which is a function of parameter $\nu_{0,363}$.
For a given  $\nu_{0,363}$ that curve can be obtained by the scaling
the argument $x$ of the polynomial $P_4(x)$:  
\begin{equation}
\nu_b(\nu_{\rm K})=C_bP_4(\nu_{\rm K}/\nu_{0,363}).
\end{equation}
The normalization of the curve is controlled by the constant $C_b$ which 
reflects the properties of the specific source.
 For 4U 1728-34 the observed frequencies are  fitted by the curve 
with $a_{\rm K}=~1.03$ (TO99) for which the constant $C_b=1$ 
with $R_0=11$ km for 1.4 solar masses.    
For 4U 0614+09 we fit the data points  by the curves with 
the parameters $\nu_{0,363} =1,~2,~3,$ (i.e. $\nu_0=$ 363 Hz, 726 Hz and 
1089 Hz)  for which  the corresponding constants $C_b=~
2,~22.8, ~55.4$ are chosen respectively. 
Using Eq (2), one can estimate mass-radius (MR) relations for a given value of
$\nu_{0,363}$.
For $\nu_{0,363}=$ 2 and 3 the radius $R_0\approx 6.4$ and 4.9  km 
respectively, for the standard NS mass, 1.4 solar masses. 
There is definitely an indication
of higher compactness for 4U 0614+09. 
It is true that there is a possibility  
 to elaborate  a much more stringent condition on the MR relation using
Eq. (2) and fits of the data in Fig. 2 which may lead to  small 
masses and radii for the star $-$ $m<1$ and $R<9$ km.
In that case such a low mass compact star can be easily explained
in terms of strange star models (e.g. Bombaci 1997). 
In fact, under constraints imposed by the TO model Li et al. 1999 pointed out 
the possibility that 4U 1728-34 is   a strange star rather than a neutron 
star. 
Thus our  arguments of the high compactness in 4U 0614+09 support 
a strange star concept. But the final conclusion on MR constraints cannot be 
worked out without the full relativistic treatment which is out of the scope 
of this paper.



According to the low panel 
in Fig. 2,  $\nu_h$ 
is not lower than 400 Hz for any value
of $\nu_{\rm K}$.
Because  
\begin{equation}
 \nu_h=[\nu_{\rm K}^2+4(\Omega/2\pi)^2]^{1/2},
\end{equation}
the rotational magnetospheric frequencies $\nu_{mag}= (\Omega/2\pi)$ should 
have an  asymptote for large radii (or small values $\nu_{\rm K}\ltorder
2\nu_{mag}$) which equals approximately 200 Hz.  
The detailed profile of $\nu_{mag}$ constructed below confirms this expectation.
\bigskip
\par
\noindent
\centerline{\bf 4. Inferred Rotational Frequency Profile of the NS 
Magnetosphere}

From the observed kHz frequencies (VS00, Table 1), $\nu_h$
and $\nu_{\rm K}$,
the profile of $\nu_{mag}=\Omega(\nu_K)/2\pi$ has been calculated 
according to formula (6) and modeled 
using  theoretically 
inferred magnetic multipole structure of differentially rotating
magnetosphere (OT99a) 
\begin{equation}
\nu_{mag}=C_0+C_1\nu_K^{4/3}+C_2\nu_K^{8/3}+C_3\nu_K^4
\end{equation}
where $C_2=2(C_1C_3)^{1/2}$. The constants
$C_0\equiv\nu^{0}_{mag}=190$ Hz, $C_1=6.52\times 10^{-2}$ Hz$^{-1/3}$,
$C_2=-1.19\times10^{-5}$ Hz$^{-5/3}$ and $C_3=0.91\times10^{-9}$
Hz$^{-3}$ have been obtained  by the least-squares fit with $\chi^2=29.8/
25$ (OT99a).              
The resulting fit is shown in the lower panel of  Fig. 3 by solid line. 
The $\nu_{mag}$ profiles  for Sco X-1
(which is very similar to those for 4U 1608-52 and 4U 1728-34)
and that for 4U 0614+09 differ in the sign of curvature:
it is negative  for the profile of Sco X-1 profile and positive for 
that of 4U 0614+09. The $\nu_{mag}$ profiles for Sco X-1, 4U 1608-52 
(as well as for 4U 1728-34) according to analytical formula 
(14) in OT99a have a maximum at large $\nu_{\rm K}$ (small radii) and also 
have a minimum at small $\nu_{\rm K}$ (large radii).  
The adjustment of the Keplerian
disk to the sub-Keplerian rotation of the NS in 4U 0614+09 
occurs at much larger distances from the central source  (3.6-7 NS radii)  
than that for 4U 1728-34 
where the adjustment radii are within 1.6-3.4 NS radii. 
Thus the blobs (formed at the adjustment radius as  result of the super$-$ 
Keplerian motion [TLM])  presumably probe the different parts of 
the magnetosphere: in the 4U 0614+09 case, it is the 
 outer part of the magnetosphere, near  the minimum of the rotational 
frequency while in the 4U 1728-34 case, it is a part which is closer  
 to the inner part, near 
the maximum. 
Thus one should  make an expansion of $\nu_{mag}$ over the Chandrasekhar 
potential $A$  near the minimum in the 
former case and near the maximum in the latter case (see OT99a for details).
 We may try to predict $\nu_{mag}$ for 
 much smaller $\nu_{\rm K}$ (around 150 Hz) than those frequencies 
$\nu_{\rm K}$  (between 400 and 850 Hz) where we have constructed  
the profile (7).  
Using VS00 data points for 8 August with 
$\nu_{\rm K}=153.4\pm 5.6$ Hz and  $\nu_h=449.4\pm 19.5$ Hz and Eq. (6),
we find $\nu_{mag}=211.\pm23.$ Hz which is in good agreement with
 the value $\nu_{mag}=235$ Hz obtained from the analytical expression, 
 Eq. (7). 
  
As shown in Fig. 3 (the upper panel) the difference in $\nu_{mag}$ 
profiles for Sco X-1 and for 4U 0614+09 results in different profiles 
for $\Delta\nu=\nu_h-\nu_{\rm K}$: namely, $\Delta\nu$  monotonically 
decreases with $\nu_{\rm K}$ for Sco X-1 but has a minimum  for 4U 0614+09.
According to VS00 $\Delta\nu=311.8\pm1.8$ Hz except one point (not included
in Fig. 3) which deviates 11 $\sigma$ from this trend.
We will return to this point in another paper. 
\bigskip
\par
\noindent
\centerline{\bf 5.  Break Frequency vs Low Lorentzian QPO Frequency Correlation}

According to solutions presented in TO99 and TOK, the break frequency $\nu_b$  
correlates 
with the low Lorentzian frequency $\nu_{V}$ (which we 
interpreted as a viscous frequency in the TO model) through power law relation 
\begin{equation}
\nu_b \approx 0.041\nu_{V}^{1.61}.
\end{equation}
It was confirmed in TOK using the data of Wijnands \& van der Klis (1999) that  
the same index approximately  is valid for the wide class of NS and 
BH objects. In Figure 4 we demonstrate that the observed correlation 
of $\nu_b$ vs $\nu_{V}$ in 4U 0614+09 is described well by the same 
power law dependence and the quality of this fit is  similar to that of 
 4U 1728-34, i.e. the Lebesgue's measure is less than 20\%.
Equations (5) and (8) give us theoretical dependences of $\nu_V$ on 
$\nu_{\rm K}$, assuming $\nu_{0,363}=3$. For August 8, 1996 the theoretical 
value $\nu_V\approx9$ Hz with 
inherited 20\% accuracy comes close to the observed $\nu_V=11.2\pm1.4$ Hz 
as given in VS00. 
 
\centerline{\bf 6. Discussion and Conclusions}

In this {\it Letter} we have verified predictions of the two-oscillator 
(TO) model for  three sources. 
The unifying characteristic of spectra for both oscillators which
share  the common boundary at the outer edge of the viscous transition 
layer is the strong dependence of all frequencies in the model 
on $\nu_{\rm K}$.  
The angle $\delta$ as a global parameter describes the
inclination of the magnetospheric equator to the equatorial plane of the 
disk.
Measured locally for different radial distances
(therefore different $\nu_K$), $\delta$ may vary considerably if 
the observed oscillations do not correspond to the predicted low
Keplerian branch. The constancy of $\delta$ shown in Figure 1 argues in favor of the TO 
model.   The parameter $\delta$ is critical in evaluating
the differences between the spectra of
different sources and we suggest 
that the higher  $\nu_L$ for 4U 0614+09 and  4U 1728-34 (than that for 
Sco X-1)  are mainly a result of a larger angles $\delta$.

We consider  the angle $\delta$ to be a fundamental geometric parameter
for neutron stars and may be  related 
to the evolution  and prehistory of these systems.
During the NS life time,  the NS is believed to spin up gaining more angular 
momentum from the companion and the  axis of magnetosphere may change 
with respect the rotational axis of the star (or the disk) (Bhattacharya 1995).
The  X-ray bursters are found in old 
population (age of order $10^9$ years) concentrated towards the galactic center 
(Lewin, van Paradijs \& Taam 1995) and thus it is not surprising
that  in these systems the angle $\delta$ between spin axis and the magnetic 
axis is small presumably as a result of long evolution.
The absence of X-ray pulsations from the LMXBs
and the fact that type I  X-ray bursters never occur in systems which 
show pulsations (Lewin, van Paradijs \& Taam 1995) probably can be explained
by the small angles $\delta$.   
In contrast, in X-ray pulsars which are mostly observed in the 
High Mass Binaries (HMXBs), $\delta$ is not small:
the mean value of $\delta$ is $27^{0}$ (Leahy 1991) and the age of these 
systems is much smaller, of order of $10^7$ years.  


The authors acknowledge discussions with
Joe Fainberg, Sergey Kuznetsov and Craig Markwardt. We are grateful to  
Steve van Straaten  for the data which enabled us to make comparisons with 
the data. We are also grateful the referee for the fruitful suggestions.

\clearpage

\begin{figure}
\caption{ $\delta-$invariant. 
Inferred angles $\delta$ between the rotational axis of 
magnetosphere and the normal to the plane of Keplerian oscillations as a 
function of low peak kHz QPO frequency for three sources: 4U 0614+09, Sco X-1,
and  4U 1702-42.  The $\delta-$invariant (eq.[3]) is calculated using the 
observational frequencies $\nu_h$, $\nu_{\rm K}$ and $\nu_L$ only.
\label{Fig.1}}
\end{figure}

\begin{figure}
\caption{ {\it Low panel:} Break frequency   of a broken power law function of
the power spectrum density  versus the frequency of the kHz QPOs 
for 4U 0614+09 (Straaten et al. 2000). Open circles are 
for the $\nu_b-$dependence versus low kHz peak and open squares are the 
$\nu_b-$dependences  versus high kHz peak. 
 The solid line is the  theoretical
curve  for break frequencies calculated using (TO99 eqs. [4-5, 9, 13]
 corresponding to the parameter $a_k=1.03$ and $\nu_0=1089$ Hz.
{\it Upper panel:} The theoretical dependences of the break frequency   
versus the low peak kHz QPOs for different values of the spin frequency
of the inner edge of accretion disk $\nu_0$. Observational points for  
4U 0614+09 (Straaten et al. 2000) {\it open circles} and for 4U 1728-34
(Ford \& van der Klis 1998) {\it solid circles}
\label{Fig.2}}
\end{figure}

\begin{figure}
\caption{{\it Low panel:} Inferred rotational frequency of the 
NS magnetosphere (solid line)  as a function of the frequency of the low kHz 
QPO peak (Eq. 7 for 4U 0614+09 and see OT99a for Sco X-1). 
For 4U 0614+09 (van Straaten et al. 2000) observational
points shown as {\it solid circles} and for Sco X-1 
(van der Klis et al. 1997) as {\it open circles}.
{\it Upper panel:} The peak difference of kHz QPO frequencies for the same 
sources. The solid lines are theoretical curves obtained using the 
above theoretical dependences of the rotational frequencies. 
\label{Fig.3}}
\end{figure}

\begin{figure}
\caption{Break frequency versus main viscous frequency
for 4U 0614+09. Solid line is the theoretical curve which is almost 
a power law with index 1.61.
\label{Fig.4}}
\end{figure}

\noindent
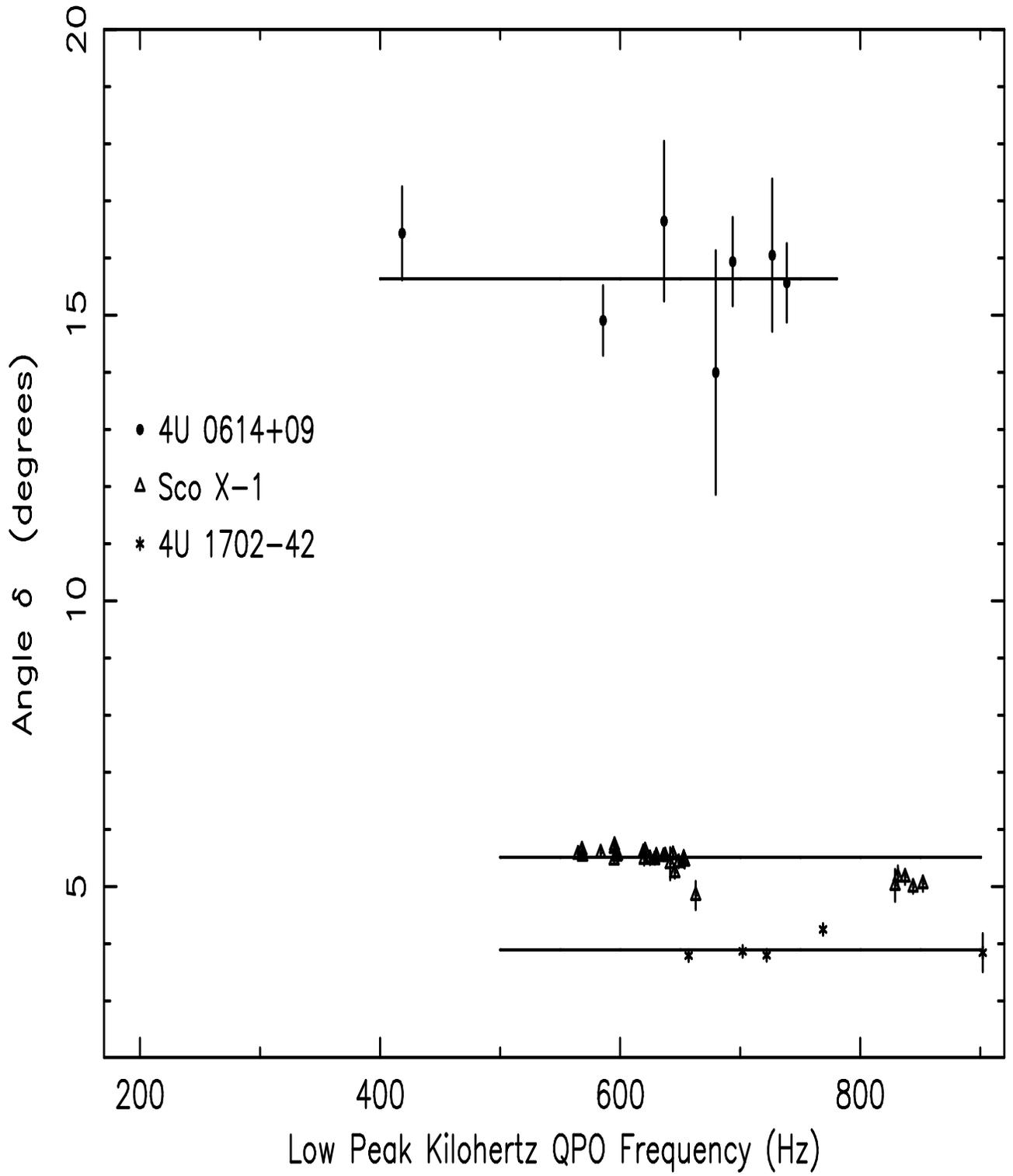
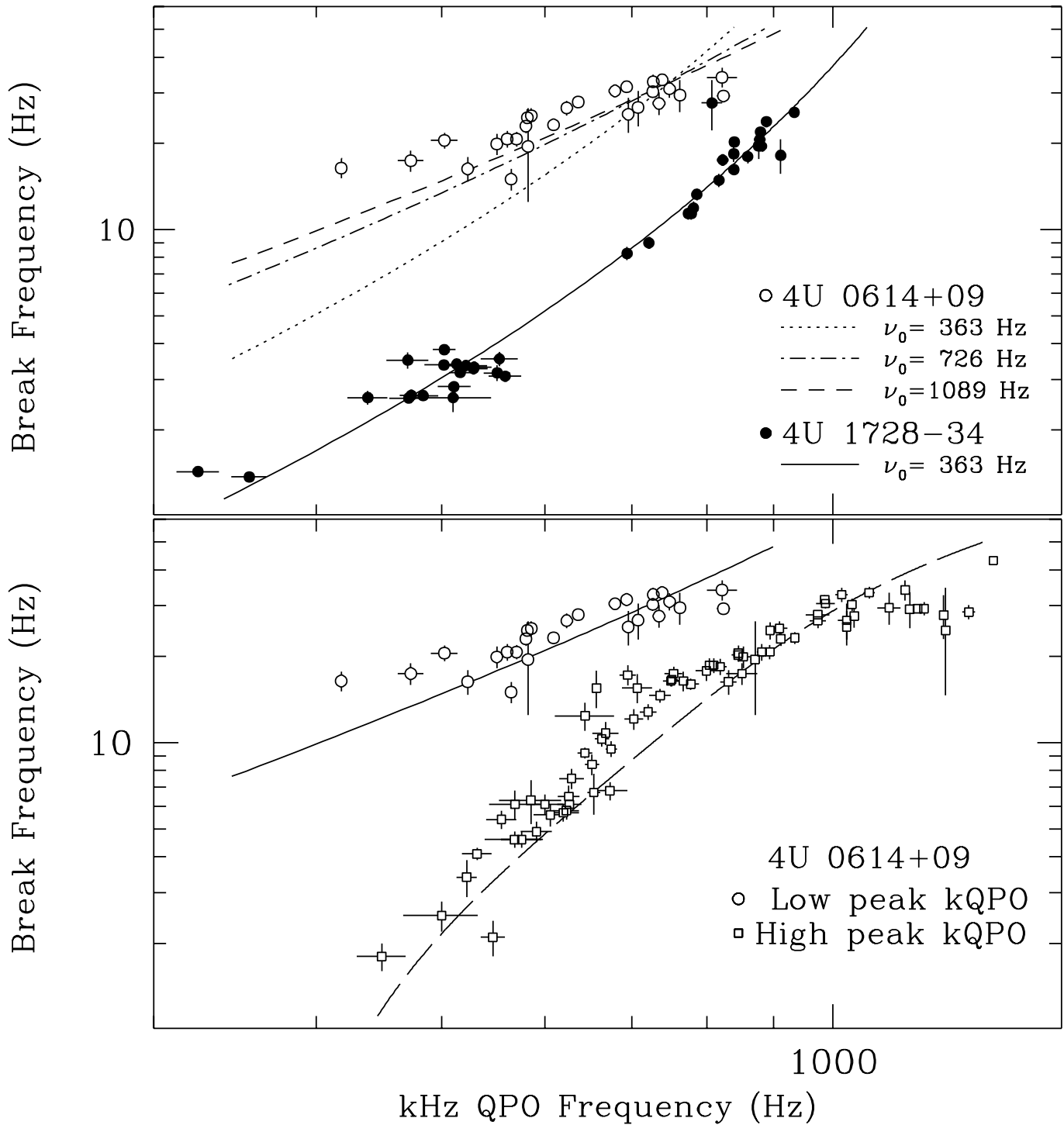
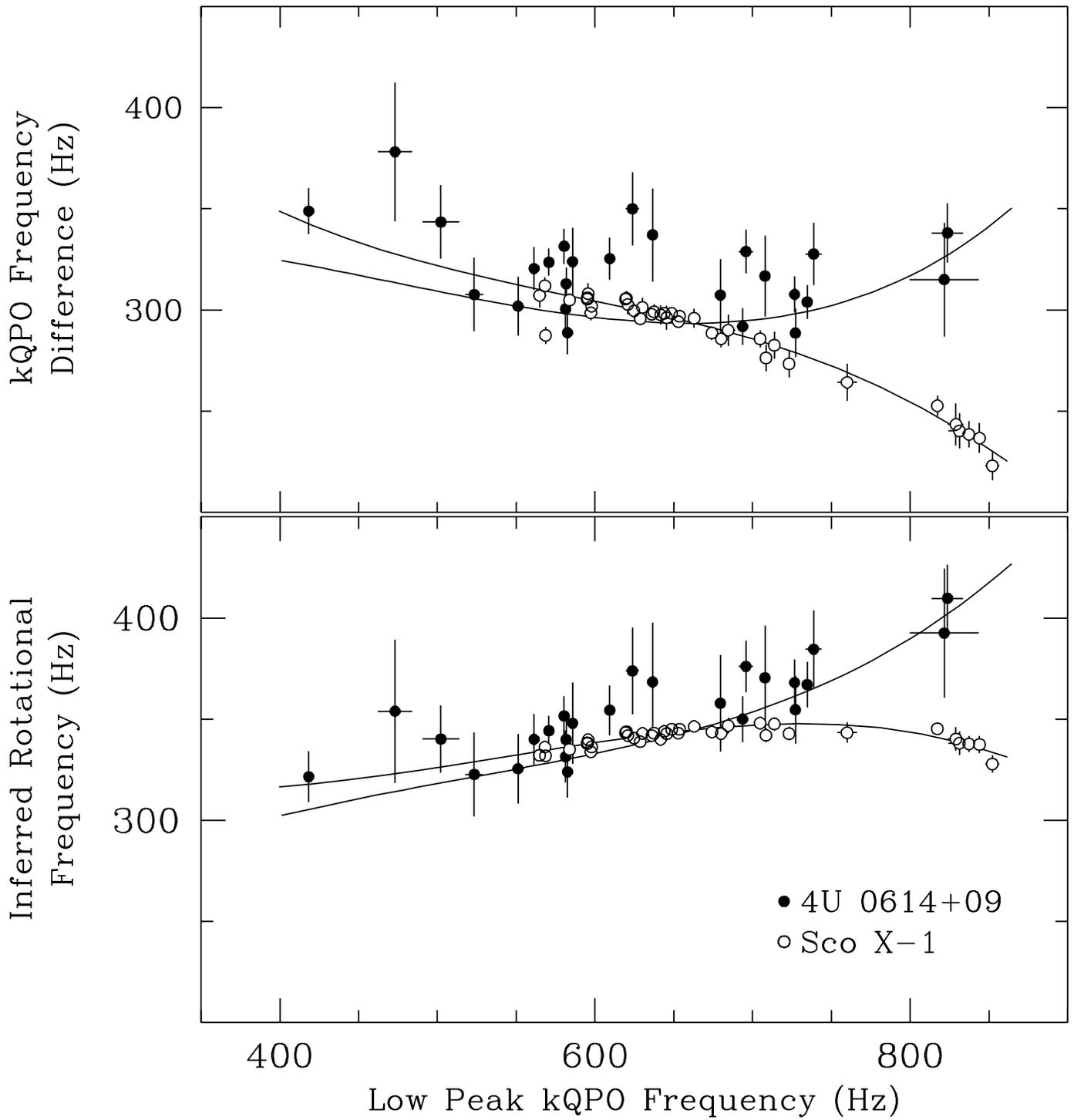
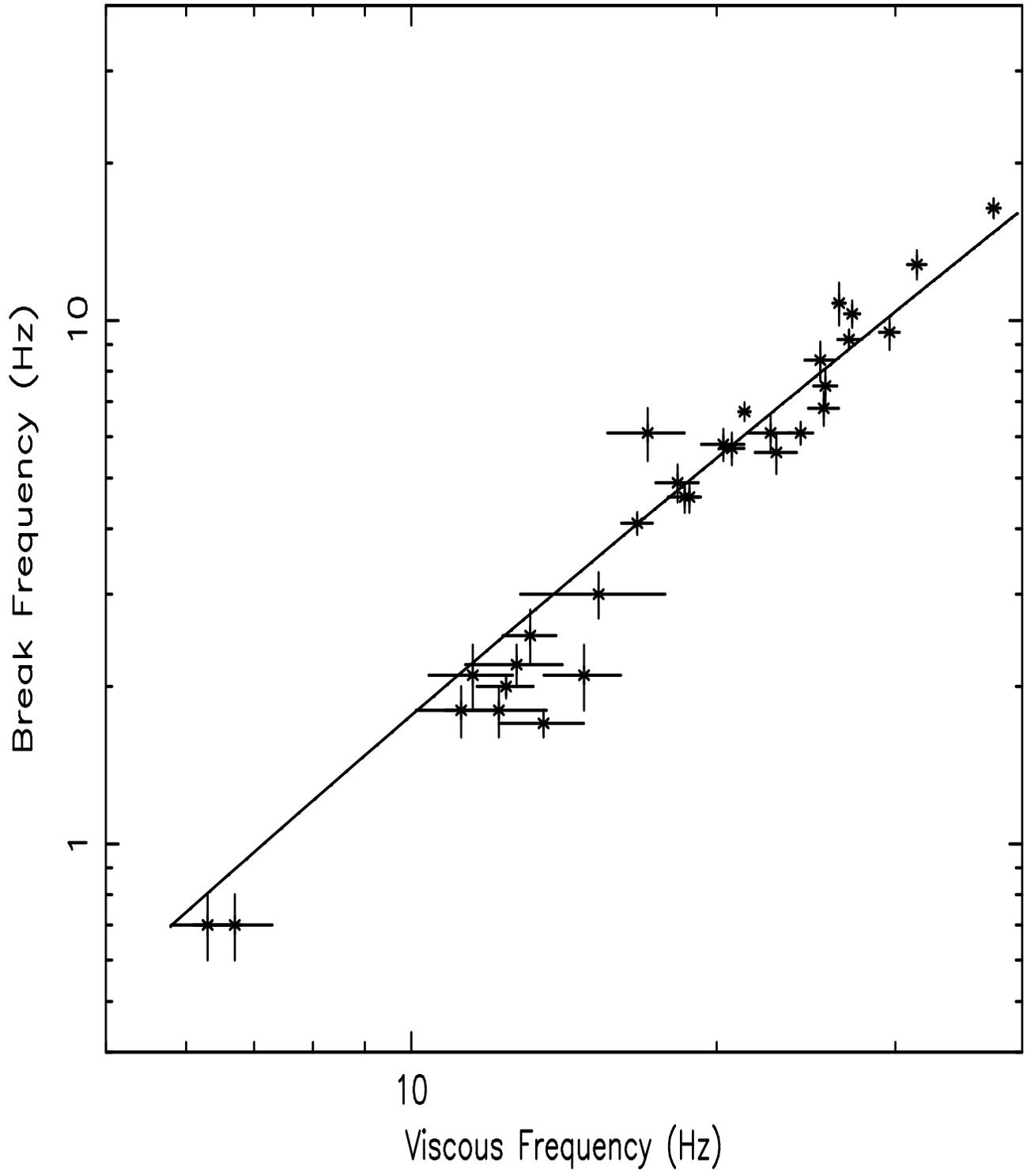
\end{document}